\documentclass[amsmath,aps,prl,reprint,groupedaddress,showpacs]{revtex4-1}

\usepackage[T1]{fontenc}

\usepackage{hyperref} 

\begin{document}

\title{Reappraisal of the limit on the variation in $\alpha$ implied by Oklo}

\author{Edward D. Davis}
\email[email: ]{edward.davis@ku.edu.kw}
\author{Leila Hamdan}
\affiliation{Physics Department, Kuwait University, P.O. Box 5969, 13060 Safat, Kuwait}

\date{\today}

\begin{abstract}
\begin{description}
\item[Background] 
A signature of many dynamical models of dark energy is that they admit variation in the fine structure constant $\alpha$ 
over cosmological time scales.
\item[Purpose]
We reconsider the analysis of the sensitivity of neutron resonance energies $E_i$ to changes in $\alpha$ with a view to 
resolving uncertainties that plague earlier treatments. 
\item[Methods]
We point out that, with more appropriate choices of nuclear parameters, the standard estimate (due to Damour and Dyson)
of the sensitivity for resonances in ${}^{150}$Sm is increased by a factor of 2.5. We go on to identify and compute 
excitation, Coulomb and deformation corrections. To this end, we use deformed Fermi density distributions fitted to the 
output of Hartee-Fock (HF) + BCS calculations (with both the SLy4 and SkM$^*$ Skyrme functionals), the energetics of the 
surface diffuseness of nuclei, and thermal properties of their deformation. We also invoke the eigenstate thermalization 
hypothesis, performing the requisite microcanonical averages with two phenomenological level densities which, via the
leptodermous expansion of the level density parameter, include the effect of increased surface diffuseness. Theoretical 
uncertainties are assessed with the \emph{inter-model} prescription of Dobaczewski et al. 
[J. Phys. G: Nucl. Part. Phys. {\bf 41}, 074001 (2014)].
\item[Results]
The corrections diminish the revised ${}^{150}$Sm sensitivity but not by more than 25\%. Subject to a weak and
testable restriction on the change in $m_q/\Lambda$ (relative to the change in $\alpha$) since the time when the Oklo 
reactors were active ($m_q$ is the average of the $\text{u}$ and $\text{d}$ current quark masses, and $\Lambda$ is the 
mass scale of quantum chromodynamics), we deduce that $|\alpha_{\text{Oklo}}-\alpha_{\text{now}}|<1.1\times 
10^{-8}\alpha_{\text{now}}$ (95\% confidence level). The corresponding bound on the present-day time variation of 
$\alpha$ is tighter than the best limit to date from atomic clock experiments.
\item[Conclusions]
The order of magnitude of our Oklo bound on changes in $\alpha$ is reliable. It is one order of magnitude lower than the 
Oklo-based bound most commonly adopted in earlier attempts to identify phenomenologically successful models of 
$\alpha$-variation. 
\end{description}
\end{abstract}

\pacs{06.20.Jr, 21.10.Ft, 24.10.Pa, 24.60.Lz}

\maketitle

Quite apart from their intrinsic metrological interest, studies of the possible variation of fundamental dimensionless 
parameters like the fine structure constant probe beyond the standard models of elementary particle physics and 
cosmology~\cite{RevModPhys.75.403, *LectNotesPhys.648, *FromVaryingCouplingsToFundamentalPhysics, 
LivingRevRelativ.14.2}. The initial application of the many-multiplet method 
to 128 Keck/HIRES quasar absorption systems suggested that, in the redshift range $0.2<z<3.7$, $\alpha$ was 
smaller than today by about 6 parts per million (ppm)~\cite{PhysRevLett.82.884, *MonNotRAstronSoc.345.609}. 
Subsequent analysis of Keck/HIRES and VLT/UVES spectra has lead to a refinement of this 
earlier claim, namely, that $\alpha$ appears to vary spatially across the sky~\cite{PhysRevLett.107.191101, 
*MonNotRAstronSoc.422.3370}. An angular dipole model of amplitude $\sim 10\,$ppm is favoured at the 
4.1$\sigma$ level over a simple monopole model in which $\alpha$ does not change across the sky but could be 
different from current laboratory values; the data also supports a dipole model with an amplitude proportional to the 
look-back time. In view of the paradigm-shifting ramifications of these findings, mined from data taken for other 
purposes, and concerns about the wavelength calibration of the HIRES and UVES spectrographs (which seem to have 
been borne out by the recent identification of long range distortions~\cite{MonNotRAstronSoc.447.446}), the UVES 
LARGE Program of dedicated observations has been initiated to check the evidence for non-zero changes in 
$\alpha$~\cite{AstronAstrophys.555.A68, *MonNotRAstronSoc.445.128}. However, the difficulties of the 
spectroscopic measurements involved demand a new generation of ultra-stable high resolution 
spectrographs~\cite{GenRelativGravit.47.1843}, two of which (PEPSI and ESPRESSO) it is envisaged will begin 
operating within a year or so.

The reconciliation of these claimed changes in $\alpha$ with stringent bounds from the
Oklo natural fission 
reactors~\cite{NuclPhysB.573.377,PhysRevC.74.024607,PhysRevC.74.064610,ModPhysLettA.27.1250232}
and single ion optical clocks~\cite{Science.319.1808, *PhysRevLett.113.210801, *PhysRevLett.113.210802}
has provoked considerable theoretical effort (work prior to 2011 is reviewed in Ref.~\cite{LivingRevRelativ.14.2}),
with several models under active development~\cite{PhysLettB.703.74, *PhysRevD.85.023514, *PhysRevD.86.043501, 
*PhysRevD.86.083517, *ResAstronAstrophys.13.1423,  *PhysRevD.89.024025, *PhysRevD.90.023017, 
PhysRevD.83.083523,*PhysRevD.90.023505,PhysRevD.88.103513,PhysRevD.89.024021}. 
Nevertheless, there is some ambivalence in the literature about the importance of the Oklo geochemical data. 
Significantly, the Oklo phenomenon is completely ignored in recent 
studies~\cite{PhysRevD.85.087301, *PhysRevD.89.083011, *PhysRevD.89.083509, *PhysRevD.90.027305, 
*PhysRevD.90.063519, *PhysLettB.743.377} of the potential of high precision measurements of the redshift dependence 
of $\alpha$ (among other quantities) to distinguish between different dynamical dark energy models and pin down 
unification scenarios. Other recent papers dismiss Oklo bounds for being ``strongly model dependent and possibly 
subject to criticism''~\cite{PhysRevD.89.024021}, ``subject to much larger theoretical and systematic uncertainties 
than \ldots spectroscopic measurements''~\cite{PhysRevD.89.123512}, and being based on a ``naive 
assumption''~\cite{JCosmolAstropartPhys.06.2014.062}, concerns no doubt based on earlier more explicit criticisms 
voiced in, for example, 
Refs.~\cite{PhysRevD.88.103513,CanJPhys.78.639, *PhysRevD.78.067304, *PhysRevD.78.103518}. In this paper, 
we aim to counteract this dismissive attitude to Oklo-based limits on $\Delta\alpha\equiv\alpha_{\text{Oklo}}-
\alpha_{\text{now}} $. (Henceforth, we shall use a subscript 0 to denote a current value, e.g.,~$\alpha_0$ is 
$\alpha_{\text{now}}$.)

Oklo data constrains shifts $\Delta E_i=E_{i\text{Oklo}}-E_{i\text{now}}$ in neutron capture resonance energies $E_i$ 
over the interval of time since the Oklo natural fission reactors were active (about $1.8\,\text{Gyr}$ ago). 
Most attention has been directed at the $n+{}^{149}{\text{Sm}}$ capture resonance nearest threshold.
Much has been made of the uncertainty associated with the modeling of the operation of the Oklo reactors and 
its impact on the values of $\Delta E_i$ extracted; in view of the resonance structure of the pertinent
neutron absorption cross-sections, shifts $\Delta E_i$ cannot exceed $50\,\text{meV}$ in 
magnitude~\cite{SovPhysUsp.20.937}, whereas, on the basis of a representative set of reactor model studies 
(see Table~\ref{tb:DEq}), we adopt, in effect, a 95\% confidence level (C.L.) upper bound on $|\Delta E_i|$ of 
$23\,\text{meV}$ for the ${}^{150}$Sm resonance: thus, other treatments of reactor dynamics cannot weaken our 
Oklo-inspired bound by more than a factor of 2. We shall also revisit the lingering issue of the ``second solution'' 
for $\Delta E_i$ encountered in some analyses of Oklo data and attempt to lend further weight to the claim of 
Ref.~\cite{NuclPhysB.573.377} that this second solution should be ignored.

As for the reduction of a bound on $\Delta E_i$ to a bound on $\Delta\alpha$, any shift $\Delta E_i$ is 
most appropriately interpreted~\cite{IntJModPhysE.23.1430007} in terms of changes in \emph{both\/} $\alpha$ and 
$X_q=m_q/\Lambda$, where $m_q$ is the average of the $\text{u}$ and $\text{d}$ current quark masses, and 
$\Lambda$ is the mass scale of quantum chromodynamics. Formally,
\begin{equation}\label{eq:DEr}
 \Delta E_i = k_q \frac{\Delta X_q}{X_{q0}} + k_\alpha \frac{\Delta\alpha}{\alpha_0},
\end{equation}
where, for the Sm resonance, it is customary to set $k_\alpha \simeq -1.1\,\text{MeV}$ (based on the work of Damour 
and Dyson~\cite{NuclPhysB.480.37}, who actually conclude that $k_\alpha < -1.1\pm 0.1\,\text{MeV}$), but,
unfortunately, $k_q$ is poorly known: on the basis of the existing estimates~\cite{PhysRevC.79.034302,Dav2014}, 
it can be argued that $|k_q|\gtrsim 10\,\text{MeV}$. (It is also conjectured~\cite{PhysRevC.79.034302} that 
$k_q$ is approximately independent of the choice of target nucleus.) For the purposes of setting an upper bound on 
$\Delta\alpha$, one can work with the related inequality
\begin{equation}\label{eq:DEineq}
| \Delta E_i | \ge \bigl| \kappa |R^\alpha_q|-1\bigr| \,  |k_\alpha| \, \frac{|\Delta\alpha|}{\alpha_0} ,
\end{equation}
where $\kappa\equiv\left| k_q/k_\alpha\right|$ and  $R^\alpha_q\equiv\left(\Delta 
X_q/X_{q0}\right)/\left(\Delta\alpha/\alpha_0\right)$. The more manageable result
\begin{equation}\label{eq:sInEq}
| \Delta E_i | \ge |k_\alpha| \,\frac{|\Delta\alpha|}{\alpha_0} 
\end{equation}
will apply whenever $R^\alpha_q=0$ or $|R^\alpha_q|\ge 2\kappa^{-1}$ [so that the factor $|\kappa |R^\alpha_q|-1|$ 
on the righthand side of Eq.~(\ref{eq:DEineq}) is greater than or equal to unity]. Given the above restriction on $|k_q|$ 
and our estimate of $k_\alpha$ [see Eq.~(\ref{eq:kalpha})], which together imply $\kappa\gtrsim 4$, Eq.~(\ref{eq:sInEq}) 
holds for ${}^{150}$Sm if $|R^\alpha_q|>\tfrac{1}{2}$.  

This criterion can be tested in any realistic model of variations in $\alpha$, which, through its dynamics, must fix a 
particular pattern of correlations between $\alpha$ and other fundamental parameters like $m_q$. In the almost 
ubiquitous
phenomenological unification scheme of 
Ref.~\cite{PhysRevD.76.023511}, which can be characterized by the free parameters $R$, $S$ and $T$, $R^\alpha_q=-
R+\tfrac{7}{9}(S+1)T$. For the canonical choices of $(R,S,T)=(30,160,\tfrac{1}{2})$~\cite{PhysRevD.84.096004} and
$(36,240,\tfrac{1}{2})$~\cite{PhysRevD.86.043529}, $R^\alpha_q= 30$ and 60, respectively. The values of $R$ and $S$ 
(with $T=\tfrac{1}{2}$) fitted~\cite{PhysLettB.724.1} to astrophysical measurements in the direction of the radio source 
PKS1413+135 and atomic clock data also suggest that 
$R^\alpha_q\sim 10$, as do the best fits of the more extensive study in Ref.~\cite{PhysRevD.89.083011}, but no 
firm conclusions can be drawn until the errors in $R$ and $S$ are substantially reduced. The situation is more clear-cut 
for the many unification scenarios of Refs.~\cite{PhysRevD.78.103518,PhysRevD.79.083533}: the smallest non-zero 
value of $|R^\alpha_q|$ (scenario 6, $\tilde{\gamma}=70$) comfortably exceeds 0.5 (by a factor of more than 3). 

In what follows, we shall assume that Eq.~(\ref{eq:sInEq}) can be used. A lower bound on $|k_\alpha|$ then suffices to 
establish an upper bound on $|\Delta\alpha|/\alpha_0$. Our focus will be on 
corrections to the Damour-Dyson estimate of a lower bound to $|k_\alpha|$. We hope to convince
even the skeptical reader that the Damour-Dyson formula is accurate to better than 25\%.

%
%


The Damour-Dyson lower bound to $|k_\alpha|$ (in our notation, 
$|k_\alpha^{\text{DD}}|$) follows from ingenuous
approximations which imply the inequality (cf.~Eq.~(43) of Ref.~\cite{NuclPhysB.480.37}, in which $Q_i$ 
is denoted by $R_1$)
\begin{equation}\label{eq:kDD}
 k_\alpha 
          < k_\alpha^{\text{DD}}\equiv - \frac{(Ze)^2}{2Q_i^3} \delta_\text{\tiny gs}\langle r^2 \rangle,
\end{equation}
where $Q_i$ is the \emph{equivalent rms radius} of the charge distribution of the compound nucleus (CN) 
state $|i\rangle$
formed by thermal neutron capture, and $\delta_\text{\tiny gs}\langle r^2 \rangle\ (>0)$ is the difference 
between the mean-square charge radii of the ground states of the daughter nucleus and of the target 
nucleus (in the case of most interest, ${}^{150}$Sm and ${}^{149}$Sm, respectively). As $k_\alpha^{\text{DD}}<0$, 
$|k_\alpha|$ is bounded from below by $|k_\alpha^{\text{DD}}|$ if the inequality in Eq.~(\ref{eq:kDD}) does, indeed, hold.

A simple matter, albeit with numerically significant consequences, concerns the choice in Ref.~\cite{NuclPhysB.480.37} of 
$Q_i=8.11\,\text{fm}$ for the ${}^{150}$Sm compound nucleus; this value is the result of a computation with a formula 
(Eq.~(50) in Ref.~\cite{Ott1989}), which contains not one but two critical transcription errors (from Eqs.~(5) and (6) in 
Ref.~\cite{AtDataNuclDataTables.14.509}): a more reasonable value, calculated directly from the measured rms charge 
radius $R_\text{ch}$~\cite{AtDataNuclDataTables.99.69} of the ground state, would be $Q_i=6.50\pm 0.01\,\text{fm}$, 
where the error arises predominantly from neglect of $\delta_\star\langle r^2\rangle$, defined in the next paragraph. [We 
estimate this contribution to the error in $Q_i$ with the expression $\delta_\star\langle r^2\rangle=-(\kappa_{\star 0}+
\kappa_\star^\prime\varepsilon)$ implicit in Eq.~(\ref{eq:defcor}), the averages of our results for $\kappa_{\star 0},
\kappa^\prime_\star$ in Table~\ref{tb:Ccoef}, and the maximal choice $\varepsilon=0.05$.] From the Sm data in Table X 
of Ref.~\cite{AtDataNuclDataTables.60.177} (which supersedes that employed in Ref.~\cite{NuclPhysB.480.37}), the 
difference in mean-square radii $\delta_\text{\tiny gs}\langle r^2 \rangle$ for the isotope pair ${}^{149,150}$Sm is 
$0.250(20)\,\text{fm}^2$, about 20\% larger than the value adopted in Ref.~\cite{NuclPhysB.480.37}. Together, these 
different parameter values imply that \emph{the value of\/} $k_\alpha^{\text{DD}}$ \emph{for Sm data should be revised 
to\/} $k_\alpha^{\text{DD}}= -2.51\pm 0.20\,\text{MeV}$.

Our basis for gauging corrections to $k_\alpha^{\text{DD}}$ is an earlier inequality [Eq.~(36)] in the analysis of
Ref.~\cite{NuclPhysB.480.37}, which relies only on the justified neglect of exchange contributions to Coulomb energies
for its validity. It reads
\begin{equation}\label{eq:mein}
 k_\alpha < \int V_i\,\delta\rho\, d^3r,
\end{equation}
where $V_i$ is the electrostatic potential of the excited CN state $|i\rangle$ and $\delta\rho=\rho^{(i)} -\rho^{(t)}$ 
is the difference between the \emph{charge\/} densities of $|i\rangle$ and the ground state $|t\rangle$ of the 
\emph{target\/} nucleus. In terms of $k_\alpha^{\text{DD}}$, the right-hand side of Eq.~(\ref{eq:mein}) is
\[
  k_\alpha^{\text{DD}} - \frac{(Ze)^2}{2Q_i^3}\,  \delta_{\star}\langle r^2\rangle \ +\
      \int\limits_{r>Q_i} {\cal V}_i\,\delta\rho\, d^3r \ +\ \int (V_i - V_u) \,\delta\rho\, d^3r,
\]
where $V_u$ is the potential of a uniformly charged sphere of radius $Q_i$ (and charge $Ze$),
\[
   {\cal V}_i  
            = \frac{Ze}{Q_i}\left[\frac{Q_i}{r} 
                 + \frac{1}{2}\left(\frac{r}{Q_i}\right)^2 - \frac{3}{2}\right] ,
\]
and $\delta_{\star}\langle r^2\rangle$ is the difference between the mean-square charge radii of the excited state 
$|i\rangle$ and the ground state of the \emph{daughter\/} nucleus. We identify the second through fourth terms above 
as the excitation, Coulomb, and deformation corrections to $k_\alpha^\text{DD}$, respectively. 

As explained in Ref.~\cite{IntJModPhysE.23.1430007},
the Coulomb correction (i.e.,~the integral involving ${\cal V}_i$) compensates for the use in Ref.~\cite{NuclPhysB.480.37} 
of the electrostatic potential appropriate to the \emph{inside\/} of a uniformly charged sphere (of radius $Q_i$) to describe
the nuclear Coulomb field throughout \emph{all\/} space. In view of the greater spatial extent of the charge distribution 
of the compound nuclear state $|i\rangle$, $\delta_{\star}\langle r^2\rangle>0$ and the density difference $\delta \rho$ 
is positive for $r>Q_i$. As a result, the excitation correction (proportional to $-\delta_{\star}\langle r^2\rangle$) is 
negative while the Coulomb correction is positive. The deformation correction is found to be
positive (see Table \ref{tb:Ccoef}). The consequent partial cancellation of these corrections proves crucial.


To compute the corrections to $k_\alpha^{\text{DD}}$, we need densities $\rho_k$ in the vicinity of the nuclear 
surface ($k=n,p,c$ for neutron, 
proton, and charge, respectively). There is not enough experimental data to permit a model independent description of 
these densities. Following the example of Refs.~\cite{PhysRevC.18.1474,NuclPhysA.316.295,NuclPhysA.364.446}, we 
adopt the deformed Fermi (DF) functions
\begin{equation}\label{eq:density}
  \rho_k = \rho_{0k}\left[1 + \exp\left( \frac{r - C_k[1+\beta_{2k} Y_{20}(\Omega ) ]}{z_k}\right) \right]^{-1} .
\end{equation}
Despite the ad hoc empirical origins of this density profile, it serves, with some modifications, as a
template for the extraction~\cite{PhysRevC.88.064327} from nuclear energy density functional theory of information 
on surface diffuseness in deformed nuclei. Evaluation of the corrections to $k_\alpha^{\text{DD}}$ involves only
charge densities, but our scheme for estimating the charge density parameters of the excited state $|i\rangle$
presupposes knowledge of neutron and proton densities separately.


Inspection of the DF density parameters obtained in experimental studies~\cite{PhysRevC.18.1474,NuclPhysA.316.295} 
of even-even Sm isotopes reveals, where comparisons are possible (${}^{150}$Sm and ${}^{152}$Sm), some systematic
inconsistencies, which suggest that uncertainties in the surface diffuseness $z_c$ and quadrupole 
deformation $\beta_{2c}$ exceed 10\,\% (and could be as much as 30\,\% or so). In the analysis of the 
${}^{149}$Sm experiment~\cite{NuclPhysA.364.446}, the value of $\beta_{2c}$ is interpolated from effective $\beta_{2c}$ 
values for ${}^{148}$Sm and ${}^{150}$Sm (relevant to Coulomb excitation). Given these difficulties with empirical
DF density parameters, we prefer to work with the theoretical $z_p$'s and $\beta_{2p}$'s inferred in 
Ref.~\cite{PhysRevC.88.064327} from HF+BCS calculations with a contact surface pairing interaction (constrained by
the findings of Ref.~\cite{PhysRevC.79.034306}) and the Skyrme functionals SkM${}^*$ and SLy4.

Consistent with the restriction to quadrupole deformation in Eq.~(\ref{eq:density}), we disregard the other
significantly smaller deformation parameters determined in Ref.~\cite{PhysRevC.88.064327}. Their effect on the 
Coulomb correction to $k_\alpha^{\text{DD}}$ is negligible. The ``surface polarization'' and the angular dependence 
of the radial diffuseness identified in Ref.~\cite{PhysRevC.88.064327} is more of a concern for the other 2 corrections, but 
these features of densities are suppressed by the angular averaging implicit in the 
calculation of volume integrals.

As Ref.~\cite{PhysRevC.88.064327} deals only with even-even nuclei, we set the $z_p$ and $\beta_{2p}$ parameters for 
${}^{149}$Sm equal to the averages of the results for ${}^{148}$Sm and ${}^{150}$Sm (the interpolation scheme of
Ref.~\cite{NuclPhysA.364.446}). We fix the values of $C_p$ and $\rho_{0p}$ for both ${}^{149}$Sm and ${}^{150}$Sm
by requiring that the proton density be normalized (to the number of protons) and that its 
second moment $\langle r^2\rangle_p$ reproduce the experimental mean-square charge radius,
calculated with the standard relation
\(
 \langle r^2 \rangle_c = \langle r^2 \rangle_p  + r_p^2  +  \tfrac{N}{Z}r_n^2
\),
where the proton rms charge radius $r_p=0.8775(51)\,\text{fm}$~\cite{RevModPhys.84.1527} and the 
neutron mean-square charge radius $r_n^2 = - 0.1161(22)\,\text{fm}^2$~\cite{ChinPhysC.38.090001}.
In fact, our values of $C_p$ and $z_p$ for ${}^{150}$Sm (see Table~\ref{tb:GSpar})
differ only very slightly from those of Ref.~\cite{PhysRevC.88.064327}, a reflection of the size of the 
deformations we have omitted. 

%
%

\begin{table}[t]
\caption{Proton and charge density parameters for ${}^{149,150}$Sm ground states.\label{tb:GSpar}}
\begin{ruledtabular}
\begin{tabular}{*{7}{c}}
Isotope        & $\rho_{0p}$      & $C_p$         & $z_p$         & $\beta_{2p}$ & $C_c$         & $z_c$ \\
               &($\text{fm}^{-3}$)& ($\text{fm}$) & ($\text{fm}$) &              & ($\text{fm}$) & ($\text{fm}$)\\ 
${}^{149}$Sm   &                  &               &               &              &               &       \\
$\text{SkM}^*$ & 0.0679           & 5.86          & 0.502         & 0.190        & 5.83          & 0.560 \\
SLy4           & 0.0679           & 5.86          & 0.503         & 0.184        & 5.83          & 0.561 \\
${}^{150}$Sm   &                  &               &               &              &               &       \\
$\text{SkM}^*$ & 0.0673           & 5.88          & 0.499         & 0.232        & 5.84          & 0.556 \\
SLy4           & 0.0671           & 5.88          & 0.501         & 0.215        & 5.85          & 0.559 \\
\end{tabular}
\end{ruledtabular}
\end{table}

The charge density parameters $C_c$ and $z_c$ in Table~\ref{tb:GSpar} are found under the reasonable assumptions that 
$\beta_{2c}=\beta_{2p}$ and $\rho_{0c}=\rho_{0p}$; the method of Appendix A in Ref.~\cite{PhysRevC.81.054309}, 
generalized to accommodate a non-zero quadrupole parameter, is used. The SkM${}^*$ charge parameters for 
${}^{150}$Sm
are very similar to the empirical parameters of Ref.~\cite{PhysRevC.18.1474}. (The SLy4 quadrupole deformations for 
${}^{149,150}$Sm agree to within 5\% with those of the finite range droplet model~\cite{AtDataNuclDataTables.59.185}.)


The complexity of CN states means that theoretical approaches can and must use statistical 
methods~\cite{RevModPhys.53.385}. Recognition of the universal character of properties of quantum chaotic 
systems (of which the compound nucleus is a prototype~\cite{RevModPhys.81.539}) broadens the scope of the 
arguments that can be brought bear to include insights deduced from studies of other more tractable many-body systems.
Central to our estimates of DF density parameters for the state $|i \rangle$ is the widely 
accepted~\cite{RevModPhys.83.863} eigenstate thermalization hypothesis (ETH)~\cite{PhysRevA.43.2046, 
*PhysRevE.50.888, *PhysRevLett.108.110601}, according to which the expectation value of a few-body observable 
$\widehat{O}$ in an individual state coincides with a microcanonical ensemble average of $\widehat{O}$, provided the 
system manifests fully developed many-body quantum chaos. More precisely, finite-size scaling 
studies~\cite{JStatMech.2011.P10028, *PhysRevE.89.042112} indicate that the relative error incurred in 
invoking the ETH decreases as $d^{-\frac{1}{2}}$, where $d$ is the dimension of the Hilbert space. As the dimensionalities
$d$ of the many-particle shell model spaces needed to describe even low-lying states in ${}^{150}$Sm (far below neutron
threshold) exceed $10^9$ by several orders of magnitude, our use of the ETH is justified to accuracies of at least 1 part in
$10^4$. Detailed nuclear shell model studies~\cite{PhysRep.276.85, *PhysRep.499.103}  in far smaller model spaces 
also confirm the appropriateness of the ETH  for quantum chaotic nuclear states. 

The ETH allows us to adapt the microcanonical ensemble 
treatment~\cite{PhysRevLett.93.132702,PhysRevC.73.014609,PhysRevC.75.017601} 
of mononuclear configurations formed in heavy ion reactions to our problem of the densities $\rho_k$ of $|i\rangle$.
The microcanonical analysis can be limited to the determination of the surface diffusenesses $z_k$ of $|i\rangle$ 
or the more convenient susceptibilities $\chi_k\equiv z_k/z_{kg}-1$, where $z_{kg}$ denotes the value 
of $z_k$ for the ${}^{150}$Sm ground state. Even at excitation energies much higher than that of $|i\rangle$, 
central densities $\rho_{0k}$ are unchanged~\cite{PhysRevC.73.014609,NuclPhysA.703.188}, which means that the 
equivalent sharp radii $R_k$, defined so that $\tfrac{4\pi}{3}R^3_k\rho_{k0}$ is equal to the volume integral of 
$\rho_k$, are also unchanged. This assumption, coupled with the fact that the quadrupole shape parameters 
$\beta_{2k}$ of $|i\rangle$ can be constrained by reference to existing
 studies~\cite{PhysRevC.62.054610,PhysRevC.63.024002,PhysRevC.63.064302} of ${}^{150}$Sm 
(see following paragraph), and the relation $R_k^3=C_k^3(1+\tfrac{3}{4\pi}\beta_{2k}^2+\pi^2 z_k^2 / C_k^2 )$, 
implies that the central radii $C_k$ can be found once the $z_k$'s are known.

The effect of excitation on the shape of ${}^{150}$Sm (and other nuclei) has been studied in 
thermal relativistic mean-field theory~\cite{PhysRevC.62.054610,PhysRevC.63.024002,PhysRevC.63.064302} 
with the versatile NL3 interaction, large model spaces (with no inert core) and a sound description of Coulomb 
interactions (all improvements on earlier investigations). For temperatures up to 
$0.75\,\text{MeV}$, corresponding to an average excitation energy in ${}^{150}$Sm exceeding the energy 
of $|i\rangle$, the reported quadrupole deformation parameter $\beta_2$ increases very slightly, due to the weakening 
of pairing correlations, as the temperature increases~\cite{PhysRevC.63.024002,PhysRevC.63.064302} ($\beta_2$ 
denotes the common value of $\beta_{2p}$ and $\beta_{2n}$). Similar behavior is observed for ${}^{164}$Er 
in a finite temperature Hartree-Fock Bogoliubov calculation~\cite{PhysRevLett.85.26}, which also uses a realistic
effective interaction (the D1S Gogny force) and a large configuration space. (The example of 
${}^{164}$Er furthermore shows that, at these temperatures, it is unnecessary to distinguish between the mean-field 
value of $\beta_2$ and its thermal average: in addition to their being numerically close, the trend in the mean-field 
value survives in the thermal average~\cite{PhysRevC.68.034327}.) 

We interpret these findings about deformation to mean that
quadrupole deformation parameters of $|i\rangle$ do not exceed their values in the ground state of ${}^{150}$Sm 
by more than 5\%, an estimate based on Fig.~1 in Ref.~\cite{PhysRevC.63.024002}. We also take the smallness of thermal
fluctuations in $\beta_2$ as justification for ignoring fluctuations in a microcanonical treatment, i.e.,~we set the
$\beta_{2k}$'s for $|i\rangle$ equal to their microcanonical ensemble averages $\overline{\beta}_{2k}$.
As the Coulomb correction to $k_\alpha^{\text{DD}}$ displays significantly more sensitivity to surface diffuseness than 
to deformation, a more careful treatment of fluctuations in the $\chi_k$'s is needed.

Structural features of ${}^{150}$Sm influencing the values of the susceptibilities $\chi_k$
include the presence of unfilled high-$n$-low-$l$ states~\cite{PhysRevC.89.064302} in the vicinity of the Fermi level 
and surface ``tidal wave'' excitations~\cite{PhysRevC.87.044333} comprising rotation-aligned octupole phonons.
Excitation of the 2.615\,MeV octupole vibration in ${}^{208}$Pb induces a change~\cite{PhysRevC.81.014602} of only 
about 1.6\% in the surface diffuseness of the relevant nuclear potential. To the extent that the octupole state in doubly 
magic ${}^{208}$Pb is typical and state-dependent changes in a nuclear potential reflect changes in the matter 
distribution, we should not expect the $\chi_k$'s  for $|i\rangle$ to be more than a few percent or so.

The energetics of small changes in surface diffuseness have been determined~\cite{PhysRevC.60.054313} within the 
self-consistent nuclear Thomas-Fermi model. Effects of deformation are ignored. For $|\chi_k|\lesssim 0.2$, 
it is found that the change in the energy of a nucleus (on Green's valley of stability) is adequately approximated 
by the quadratic form
\[
 \Delta E = \tfrac{1}{2}(18.63\,\mathrm{MeV})A^\frac{2}{3}\left( \phi_1\chi_n^2 
            - 2 \phi_2 \chi_n \chi_p
            + \phi_3 \chi_p^2\right) , 
\]
where the coefficients $\phi_i$ are presented in Table I of Ref.~\cite{PhysRevC.60.054313} as cubic fits in
$x=(Z/100)^\frac{1}{3}$ to numerical results.

We require that the most probable values of the $\chi_k$'s at any CN excitation energy $E^*$ maximize the entropy $S$. 
As regards the choice of the level density (the logarithm of which yields $S$), 
it has been suggested~\cite{PhysRevC.86.044322} that a compo{\-}\hbox{site} Gilbert-Cameron (CGC) formula, with
a larger energy shift and a multiplicative enhancement in the Fermi gas regime, encapsulates qualitatively the 
influence of shell effects, collective excitations and pairing correlations on the level density.
In lieu of specific information on the magnitude of these modifications for ${}^{150}$Sm, we employ both
the back-shifted Fermi gas (BSFG) and the constant temperature (CT) formulas, with ${}^{150}$Sm parameters 
(appropriate to $E^*<10\,\text{MeV}$) taken from Table II in Ref.~\cite{PhysRevC.80.054310}. Together, the 
BSFG and CT models should bracket the range of behavior manifested by a modified CGC formula.
 
For each of these models, we follow Ref.~\cite{PhysRevC.75.017601} and subtract from $E^*$
 the diffuseness expansion energy $\Delta E$: e.g., we set the BSFG entropy
\(
 S_\text{BSFG} = 2\sqrt{a(E^* - \Delta E -  E_1)},
\)
where $E_1$ is the BSFG energy shift of Ref.~\cite{PhysRevC.80.054310}. We also incorporate the impact
of increasing surface diffuseness on the level density parameter $a$ in the BSFG model. Guided by the
structure of the standard leptodermous expansion of $a=a_n+a_p$~\cite{NuclPhysA.372.141},
we demand that
\begin{equation}\label{eq:aBSFG}
  a = \sum_k a_{0k} \left[ 1 + \kappa \frac{z_{kg}}{R_k} ( 1 + \chi_k ) \right] ,
\end{equation}
where the strength $\kappa$ of the surface terms is fixed so that the systematic 
$A$-dependence~\cite{NuclDataSheets.110.3107} of $a_\text{BSFG}$ is reproduced, i.e.,
\(
 \kappa \sum_k a_{0k} \tfrac{z_{kg}}{R_k} \left/ \sum_k a_{0k} \right. = (\beta/\alpha)A^{-\frac{1}{3}} 
\)
with $\alpha$ and $\beta$ taken from Eq.~(61) in Ref.~\cite{NuclDataSheets.110.3107}. The $a_{0k}$'s are 
chosen so that $a_{0n}/a_{0p}=(N/Z)(\rho_{0p}/\rho_{0n})^\frac{2}{3}$ 
(cf.~Eqs.~(13) and (14) in Ref.~\cite{NuclPhysA.372.141}), 
and Eq.~(\ref{eq:aBSFG}) reduces to the value of $a_\text{BSFG}$ in Ref.~\cite{PhysRevC.80.054310} when 
the $\chi_k$'s are zero.
[Instead of a single factor $\kappa$, Eq.~(\ref{eq:aBSFG}) should, in principle, contain two deformation dependent 
factors $\kappa_k$, approximately proportional to $1+\beta_k^2/\pi$, but the ratio $\kappa_p/\kappa_n$ differs from
unity by less than 0.5\%.]
Finally, we employ the relation~\cite{PhysRevC.80.054310}  between $a_\text{BSFG}$ and the temperature parameter in 
the CT model to substitute this temperature parameter in the denominator of $S_\text{CT}$ by $5.164 a^{-0.791}$, 
where $a$ is given by Eq.~(\ref{eq:aBSFG}). As the lefthand half of Table~\ref{tb:CHIp} illustrates, our values of the 
$a_{0k}$'s and $\kappa$ are insensitive to whether we adopt the SkM${}^*$ or the SLy4 ground state densities of 
Ref.~\cite{PhysRevC.88.064327}.

\begin{table}[t]
\caption{Parameters of $a$ in Eq.~(\ref{eq:aBSFG}) ($a_{0k}$'s in 
$\text{MeV}^{-1}$) and of the Gaussian $\chi_p$-distributions in the BSFG and CT models. \label{tb:CHIp}}
\begin{ruledtabular}
\begin{tabular}{l*{3}{c}@{\hspace*{2em}}r*{2}{c}}
           &$a_{0p}$      &$a_{0n}$      &$\kappa$&      & $\mu$   & $\sigma$ \\
$\text{SkM}^*$& 5.44      & 6.29         & 5.72   & BSFG & 0.00806 & 0.0541   \\
SLy4       &    5.44      & 6.28         & 5.68   & CT   & 0.00696 & 0.0494   \\
\end{tabular}
\end{ruledtabular}
\end{table}

For each choice of $S$, the related microcanonical probability distribution function for the $\chi_k$'s is well 
represented about its maximum by a bivariate Gaussian. The means $\mu$ and standard deviations $\sigma$
of the associated Gaussian marginal distributions for $\chi_p$, listed in Table~\ref{tb:CHIp}, are the same (to 3 
significant figures) for our two sets of $(a_{0p},a_{0n},\kappa )$.


We can now discuss estimates of the corrections to $k_\alpha^{\text{DD}}$. We begin with the excitation correction, 
proportional to $\delta_\star\langle r^2 \rangle$. 
The relation between $\langle r^2\rangle=\langle r^2\rangle_c$ and $\langle r^2\rangle_p$ quoted 
above allows us to set $\delta_\star\langle r^2\rangle=\delta_\star\langle r^2\rangle_p$.
There are two contributions to
$\delta_\star\langle r^2\rangle_p$, one proportional to $\delta_\star\beta_{2p}^2$, the other proportional to 
$\delta_\star z_p^2$. 
In line with our earlier assumptions about $\beta_{2p}$, we approximate 
$\delta_\star\beta^2_{2p}$ as $\overline{\beta}_{2p}{\hspace*{-1.5ex}{}^2}\hspace*{0.5ex}-\beta_{2pg}^2$, 
where $\beta_{2pg}$ is the value of $\beta_{2p}$ for the ${}^{150}$Sm ground state. We calculate $\delta_\star z_p^2$ 
by averaging over the Gaussian distribution for $\chi_p$. Thus, 
\begin{equation}\label{eq:dstar}
 \delta_\star\langle r^2 \rangle = \tfrac{3}{4\pi} R_p^2  
                               \left[ \overline{\beta}_{2p}^{\,2} - \beta_{2pg}^2 \right]   
                                + \pi^2 z^2_{pg} (2\mu+\mu^2+\sigma^2) .
\end{equation}
For the small range of $\overline{\beta}_{2p}$ values we admit (from $\beta_{2pg}$ to $1.05\beta_{2pg}$), the 
excitation correction is excellently approximated by a linear 
function of $\varepsilon=\overline{\beta}_{2p}/\beta_{2pg}-1$, which it is natural to write as
\begin{equation}\label{eq:defcor}
  + \frac{(Ze)^2}{2Q_i^3}(\kappa_{\star 0}+\kappa_\star^\prime \varepsilon ) .
\end{equation}
The constant coefficients $\kappa_{\star 0}$ and $\kappa_\star^\prime$ are given in Table~\ref{tb:Ccoef}. 

\begin{table*}
\caption{Individual correction coefficients $\kappa_{i0},\kappa_i^\prime$ ($i=c,\star,d$) for ${}^{150}$Sm data
and net correction coefficients $\kappa_0,\kappa^\prime$ for ${}^{150}$Sm, ${}^{156}$Gd and ${}^{158}$Gd data.
(All coefficients are in units of fm${}^2$.)
\label{tb:Ccoef}}
\begin{ruledtabular}
\begin{tabular}{lc@{\hspace*{1em}}*{6}{c}@{\hspace*{1.5em}}*{2}{c}@{\hspace*{1.5em}}*{2}{c}@{\hspace*{1.5em}}*{2}{c}}
              &      &\multicolumn{6}{c}{${}^{150}$Sm} &\multicolumn{2}{c}{${}^{150}$Sm} 
                                                       & \multicolumn{2}{c}{${}^{156}$Gd} & \multicolumn{2}{c}{${}^{158}$Gd} \\
              &      &$\kappa_{c0}$     &$\kappa_c^\prime$    &
                      $\kappa_{\star 0}$&$\kappa_\star^\prime$& 
                      $\kappa_{d0}$     &$\kappa_d^\prime$    &
                      $\kappa_{0}$      &$\kappa^\prime$      &
                      $\kappa_{0}$      &$\kappa^\prime$      & 
                      $\kappa_{ 0}$     &$\kappa^\prime$      \\
$\text{SkM}^*$& BSFG & 0.0320           & 0.119               &
                     $-0.0469$          & $-0.964$            & 
                       0.0672           & 0.543               &
                       0.052            & $-0.30$             & 
                     $-0.005$           & $-0.68$             &
                     $-0.018$           & $-0.75$             \\
              &   CT & 0.0302           &   "                 &
                     $-0.0403$          &   "                 & 
                       0.0669           &   "                 &
                       0.057            &   "                 & 
         \hphantom{$-$}0.000            &   "                 &
                     $-0.014$           &   "                 \\
SLy4          & BSFG & 0.0305           & 0.101               &
                     $-0.0474$          & $-0.833$            & 
                       0.0400           & 0.452               &
                       0.023            & $-0.28$             & 
         \hphantom{$-$}0.012            & $-0.63$             &
                     $-0.016$           & $-0.72$             \\
              &   CT & 0.0287           &   "                 &
                     $-0.0407$          &   "                 & 
                       0.0397           &   "                 &
                       0.028            &   "                 & 
         \hphantom{$-$}0.016            &   "                 &
                     $-0.011$           &   "                 \\
\end{tabular}
\end{ruledtabular}
\end{table*}

Our calculations reveal that the Coulomb and deformation corrections can also be regarded as linear functions of
$\varepsilon$. For ease of comparison, we adopt parametrizations of the same form
as in Eq.~(\ref{eq:defcor}) with $\kappa_{\star 0},\, \kappa_\star^\prime$ replaced by coefficients
$\kappa_{i 0},\, \kappa_i^\prime$ ($i=c,d$), again tabulated in Table~\ref{tb:Ccoef}. [Here, $c\, (d)$ denotes 
the Coulomb (deformation) correction.]

Concerning the Coulomb correction to $k_\alpha^{\text{DD}}$, the integral $I$ of the product of the function ${\cal V}_i$ 
with a DF charge density $\rho^{(\alpha)}$ over the volume outside a sphere of radius $Q_i$ can be reduced to the angular 
average of a linear combination of complete Fermi-Dirac integrals ${\cal F}_k(\eta)$~\cite{ApplSciResB.6.225}:
\[
    I = \frac{(Ze)^2}{3Q_i} \left(\frac{3z_{c\alpha}}{R_c}\right)^{\hspace*{-0.5ex}3}  
     \sum\limits_{k=0}^2 \binom{2}{k} 
                         \Bigl\langle {\cal F}_{2+k}\left(-\eta_\alpha\right)\Bigr\rangle_\Omega 
                         \left(\frac{2z_{c\alpha}}{Q_i}\right)^{\hspace*{-0.5ex}k} ,
\]
where $\eta_\alpha\equiv \left(Q_i-C_{c\alpha}[1+\beta_{2c\alpha}Y_{20}(\Omega)]\right)/z_{c\alpha}$ and 
$\langle\cdots\rangle_\Omega$ denotes the average over all solid angles $\Omega$. The Coulomb correction is
$\delta_\text{\tiny gs} I +\delta_\star I$. The $\delta_\text{\tiny gs} I$-term is computed with the charge 
distribution parameters of Table~\ref{tb:GSpar}. In the $\delta_{\star}I$-term, which entails averaging over the 
$\chi_c$-distribution, we substitute $\beta_{2ci}$ by the range of $\overline{\beta}_{2p}$-values considered above
and assume that the $\chi_c$-distribution can be approximated by that of $\chi_p$. The $\chi_c$-dependence of the
central radius $C_{ci}$ is also taken into account.

Calculation of the deformation correction is facilitated by multipole expansions in even $l$ spherical harmonics 
$Y_{l0}$. For convergence to 3 significant figures, the $l=0,2,$ and 4 terms suffice. The quadrupole contribution to 
the deformation correction is dominant. The $\chi_c$ and $\beta_{2ci}$ dependence is dealt with in the same way 
as that of the Coulomb correction.
 
Together, our results for ${}^{150}$Sm in Table \ref{tb:Ccoef} imply that, depending on the actual value of $\varepsilon$ 
and the choice of model (SkM${}^*$+BSFG, etc.), the excitation and deformation corrections are separately somewhere 
between 15\% and 40\% of the revised value of $k_\alpha^\text{DD}$, and the Coulomb correction is about 10-to-15\%. 
However, because of the partial cancellation of these corrections, the \emph{net\/} correction $k_\alpha^\text{corr}$ is 
relatively modest: $k_\alpha^\text{corr}$ is always less than 25\% of the revised value of $k_\alpha^\text{DD}$. Following 
the \emph{inter-model\/} strategy of Ref.~\cite{JPhysG-NuclPartPhys.41.074001} for the estimation of the error associated 
with the use of models, we compute the net correction $k_\alpha^\text{corr}$  for the two extreme values of $\varepsilon$ 
(i.e., 0 and 0.05) in all four of the models in Table~\ref{tb:Ccoef}, and employ the average of these values and their standard 
deviation as, respectively, our best estimate for $k_\alpha^\text{corr}$ and the related uncertainty:
\begin{equation}\label{eq:kalphaCORR}
k_\alpha^\text{corr} = 0.33\pm 0.16\,\text{MeV} .
\end{equation}
(More robust statistics~\cite{AmStat.57.233}, namely, the median and the median average deviation from the median, yield 
almost the same numerical result: $k_\alpha^\text{corr} = 0.33\pm 0.14\,\text{MeV}$.)

The preceding analysis implies that, for the ${}^{150}$Sm resonance of interest, 
\begin{equation}\label{eq:kalpha}
k_\alpha< k_\alpha^\text{Bd} \equiv k_\alpha^\text{DD} + k_\alpha^\text{corr} = -2.18\pm 0.26\,\text{MeV},
\end{equation}
where the errors in $k_\alpha^\text{DD}$ and $k_\alpha^\text{corr}$ have been added in quadrature. 

The generic character of our arguments about corrections to $k_\alpha^{\text{DD}}$ means that they can be
adapted to other complex nuclei, in particular, the well-deformed ${}^{156}$Gd and ${}^{158}$Gd isotopes 
considered in Ref.~\cite{NuclPhysB.573.377}. 
Using the Gd data in Refs.~\cite{AtDataNuclDataTables.99.69,AtDataNuclDataTables.60.177}, $k_\alpha^{\text{DD}}
= -1.1\pm 0.1\,\text{MeV}\; (-1.3\pm 0.2\,\text{MeV})$ for thermal neutron capture by 
${}^{155}\text{Gd}\; ({}^{157}\text{Gd})$. 
The differences in $k_\alpha^{\text{DD}}$-values are primarily a consequence of the variable extent of odd-even 
staggering in mean-square radii, an effect which cannot be reproduced 
by an approximation~\cite{PhysRevD.66.045022} based on the Coulomb term in the Bethe-Weizs\"{a}cker mass formula.
The only aspect of our treatment of corrections which requires modification is the choice of 
maximal value for $\varepsilon$. 
As the ground state quadrupole deformations $\beta_2$ of ${}^{156}$Gd and ${}^{158}$Gd are very close to that of 
${}^{164}$Er, we appeal to the thermal behavior of $\beta_2$ for ${}^{164}$Er in Fig.~2(a) of 
Ref.~\cite{PhysRevC.68.034327} 
to constrain $\varepsilon$ to the interval between 0 and 0.03.
The net corrections to the $k_\alpha^{\text{DD}}$-values above are then, parallelling the earlier Sm analysis, 
$-0.04\pm 0.13\,\text{MeV}$ and $-0.26\pm 0.11\,\text{MeV}$ for ${}^{156}$Gd and ${}^{158}$Gd, respectively. 
The corresponding limits on $k_\alpha$ are $k_\alpha< -1.1\pm 0.2\,\text{MeV}$ for ${}^{156}$Gd and
$k_\alpha< -1.6\pm 0.2\,\text{MeV}$ for ${}^{158}$Gd.
 
These results for Gd isotopes have a bearing on the resolution of the ``second solution'' 
problem~\cite{NuclPhysB.573.377}. From the comparison~\cite{EarthPlanetSciLett.30.94} of measured rare earth 
isotope abundances in Oklo sample KN50-3548
with calculations based on present-day neutron absorption cross sections, it can be inferred~\cite{SovPhysUsp.20.937} that
shifts $\Delta E_i$ in resonances must be less than $50\,\text{meV}$ in magnitude. Nevertheless, the 
results~\cite{NuclPhysB.480.37} of Damour and Dyson for Oklo samarium data alone can be 
interpreted~\cite{NuclPhysB.573.377} to mean that $\Delta E_i$ lies in either a ``right branch'' 
$\Delta E_i=46\pm 22\,\text{meV}$, compatible with zero, or a ``left branch'' $\Delta E_i=-94\pm 13\,\text{meV}$, 
inconsistent with the $50\,\text{meV}$ bound on $|\Delta E_i|$. Subsequently, for Oklo RZ10 and RZ13 samples, a 
similar non-null solution was again found (in addition to a null solution) by Fujii et al.~\cite{NuclPhysB.573.377} 
($\Delta E_i=-97\pm 8\,\text{meV}$) and Gould et al.~\cite{PhysRevC.74.024607} 
($\Delta E_i=-90.8\pm 5.6\,\text{meV}$).

Exciting as the prospect of a non-zero result for $\Delta E_i$ may be, we are of the opinion that
the intervals which do not overlap with zero are an artifact, ultimately, of the symmetry of a Breit-Wigner  
absorption cross section about the resonance energy. If there is a (physical) energy interval to one side of a resonance
which can be associated with a particular range of effective capture cross section values, then
there will necessarily also be an unphysical interval on the other side of the resonance. 

\begin{table}[b]
\caption{1$\sigma$ intervals for $\Delta E_i$ and their sources. \label{tb:DEq}}
\begin{ruledtabular}
\begin{tabular}{*{5}{c}}
       & $E_i$        & $k_\alpha^{\text{DD}} \Delta\alpha/\alpha_0$
                                                         & $\Delta E_i$             &                          \\
       & (\text{meV}) & (\text{meV})                     & (\text{meV})             & Ref.                     \\ \hline
       &              &                                  & \hphantom{$-$1}$4\pm 16$ & \cite{NuclPhysB.573.377} \\
$\text{n}+{}^{149}\text{Sm}$                                  
       & 97.3         & 2.2                              &  $ 7.2 \pm 9.4$  & \cite{PhysRevC.74.024607} \\
       &              &                                  & $1.9 \pm  4.5$  & \cite{ModPhysLettA.27.1250232} \\ \hline
$\text{n}+{}^{155}\text{Gd}$
       & 26.8         & 0.9                              & $-8.5  \pm 17.5$    & \cite{NuclPhysB.573.377} \\ \hline
$\text{n}+{}^{157}\text{Gd}$
       & 31.4         & 1.1                              & $-8.5  \pm 17.5$    & \cite{NuclPhysB.573.377} \\
\end{tabular}
\end{ruledtabular}
\end{table}

An attempt to reconcile Oklo data on thermal neutron capture by all three of the isotopes ${}^{149}$Sm, 
${}^{155}$Gd, and ${}^{157}$Gd reinforces this point. Generalizing the analysis of Ref.~\cite{NuclPhysB.573.377} 
to include the $\Delta X_q$-term in Eq.~(\ref{eq:DEr}), one expects that $\Delta E_i^q\equiv\Delta E_i - 
k_\alpha \Delta\alpha/\alpha_0$ should be \emph{approximately the same for all nuclei}~\cite{PhysRevC.79.034302}. 
For the sake of argument, we calculate the $\Delta\alpha$-contribution to $\Delta E_i^q$ using a quasar-based 
estimate of $\Delta\alpha\simeq -8.6\times 10^{-10}\alpha_{\text{now}}$~\cite{EPLett.97.20006} and the values above 
of $k_\alpha^{\text{DD}}$. (In the quasar-based estimate, $\Delta\alpha$ is attributed to the motion of
our local galaxy cluster relative to the Australian dipole in the time since the Oklo reactors were active.)
The corresponding values of the $\Delta\alpha$-term, presented in Table~\ref{tb:DEq}, are an order of 
magnitude smaller than the $\Delta E_i$'s used in Ref.~\cite{NuclPhysB.573.377} (also given in Table~\ref{tb:DEq}). 
Hence, the conclusions in Ref.~\cite{NuclPhysB.573.377} about $\Delta E_i$'s will apply to the $\Delta E_i^q$'s: 
if one admits the presence of post-reactor contamination in the Gd data (at the 3-to-4\% level), then one can 
isolate $\Delta E_i^q$-intervals for all three nuclei which are approximately the same in as much as they all
overlap zero, whereas the unphysical interval is \emph{negative\/} for the case of Sm and \emph{positive\/} for 
the Gd isotopes. This pattern will continue to apply if the actual values of the $\Delta\alpha$-term are used 
provided, of course, that they do not differ substantially from the choices in Table~\ref{tb:DEq}.

Despite the many uncertainties to which the analysis of Oklo data is subject, the different Sm results
for $\Delta E_i$ in Table~\ref{tb:DEq} agree to within a factor of 2 with the result of Ref.~\cite{PhysRevC.74.024607}, 
which we adopt. If we combine the bound of Ref.~\cite{PhysRevC.74.024607} on $\Delta E_i$ with our restriction in 
Eq.~(\ref{eq:kalpha}) on $k_\alpha$, and simplify the distribution of the ratio $-\Delta E_r/k_\alpha^\text{Bd}$
along the lines of Sec. 4 of Ref.~\cite{JStatSoft.16.04}, we deduce the upper bound at 95\% C.L. of
\begin{equation}\label{eq:DeltaAlpha}
  \frac{|\Delta\alpha|}{\alpha_0} < 1.1\times 10^{-8} ,
\end{equation}
which is comparable to the Oklo-based limits listed in Refs.~\cite{IntJModPhysE.23.1430007} 
and~\cite{ProgTheorPhys.126.993}, but on a sounder footing. The quasar-based prediction in 
Ref.~\cite{EPLett.97.20006} of $|\Delta\alpha|/\alpha_0$ is compatible with Eq.~(\ref{eq:DeltaAlpha}).

Assuming a linear time dependence for $\alpha$ over the last 1.8 billion years, Eq.~(\ref{eq:DeltaAlpha}) implies that
the present-day time variation of $\alpha$ is subject to the (95\% C.L.) bound
\[
  \left( \frac{1}{\alpha}\left|\frac{d\alpha}{dt}\right| \right)_0 < 0.61\times 10^{-17}\,\text{yr}^{-1},
\]
which is an improvement on the best limit from atomic clock experiments~\cite{PhysRevLett.113.210801}.


In this paper, we have been at pains to demonstrate that the order of magnitude of the bound in Eq.~(\ref{eq:DeltaAlpha}) 
is reliable. We believe that neglect of the Oklo-based bound on $\Delta\alpha$ is unfortunate. It provides a restrictive 
low-$z$ datum which can help to select from the current plurality of theoretical models admitting variations in $\alpha$, 
those which are phenomenologically acceptable. Most of the model studies which have included the Oklo limit on 
$\Delta\alpha$ in their analysis, have 
been content to invoke the result of $|\Delta\alpha|/\alpha_0\lesssim 10^{-7}$ to be found in Damour and Dyson's 
seminal paper~\cite{NuclPhysB.480.37}. It would be interesting to see how previous conclusions are revised if a bound on 
$|\Delta\alpha|/\alpha_0$ of the order of $10^{-8}$ is adopted.
Models \cite{PhysRevD.83.083523,PhysRevD.88.041504, *PhysRevD.90.024001,PhysRep.568.1} 
which naturally suppress the variation of $\alpha$ in the presence of matter may well be preferred to the exclusion of all
others. It should also be instructive to consider the impact of this bound on feasibility studies
pertaining to the ambitious program of astrophysical measurements of the redshift dependence of parameters
like $\alpha$ put forward
in Ref.~\cite{PhysRevD.74.083508} and reviewed recently in Ref.~\cite{GenRelativGravit.47.1843}. 

\begin{acknowledgments}
E.D.D. would like to thank the Physics Department at NCSU, Raleigh for its hospitality during the early stages 
of this work and for continued remote access to its Library facilities. He also thanks Chris Gould for his comments on an
earlier version of this manuscript.
\end{acknowledgments}

\bibliography{oklonew}

\end{document}